\documentclass[runningheads]{llncs}
\usepackage{graphicx}
\newcommand{\llm}{LLM}

\newcommand{\youtube}{YouTube}
\newcommand{\relatedcourse}{baseline courses}
\newcommand{\groundcourse}{ground truth courses}
\newcommand{\gptthree}{GPT-3.5}
\newcommand{\gptfour}{GPT-4}
\newcommand{\onlinelearning}{Online Learning}
\newcommand{\bert}{\texttt{BERTScore}}

\usepackage{booktabs}
\usepackage{multirow}
\usepackage{hyperref}
\usepackage{amsmath}
\usepackage{cleveref}
\usepackage{pgf}
\usepackage{soul}
\usepackage{enumitem}
\renewcommand\st[1]{\ifhmode\unskip\fi}%
\usepackage{csquotes}
\usepackage{xcolor}
\usepackage[%
]{pdfcomment}

\everymath=\expandafter{\the\everymath\displaystyle}
\everymath=\expandafter{\the\everymath\displaystyle}

\usepackage{ifthen}
\usepackage{ulem}
\usepackage{xcolor}
\newboolean{mocommentenabled}
\newboolean{onlysuggestedtext}
\newcommand{\enablemocomment}{\setboolean{mocommentenabled}{true}}

\newcommand{\mocomment}[2]{%
  \ifthenelse{\boolean{mocommentenabled}}{%
    \ifthenelse{\boolean{onlysuggestedtext}}{%
      \ifthenelse{\equal{#2}{}}{#1}{#2}%
    }{%
      \ifthenelse{\equal{#2}{}}{#1}{\sout{#1}\textcolor{blue}{#2}}%
    }%
  }{#1}%
}
\enablemocomment
\usepackage{siunitx}

\usepackage{etoolbox}
\begin{document}
\title{Beyond Search Engines: Can Large Language Models Improve Curriculum Development?}

%
\author{Mohammad Moein\inst{1} \and
Mohammadreza Molavi Hajiagha\inst{1} \and
Abdolali Faraji\inst{1} \and Mohammadreza Tavakoli\inst{1} \and Gábor Kismihók\inst{1}}
\authorrunning{M. Moein et al.}
%\institute{Princeton University, Princeton NJ 08544, USA \and
%Springer Heidelberg, Tiergartenstr. 17, 69121 Heidelberg, Germany
\institute{Technical Information Library, Hannover, Germany\\
\email{\{mohammad.moein\}@tib.de}}

\maketitle              % typeset the header of the contribution
\begin{abstract}
    While \onlinelearning{} is growing and becoming widespread, the associated curricula often suffer from a lack of coverage and outdated content. In this regard, a key question is how to dynamically define the topics that must be covered to thoroughly learn a subject (e.g., a course). Large Language Models (\llm{}s) are considered candidates that can be used to address curriculum development challenges. Therefore, we developed a framework and a novel dataset, built on \youtube{}, to evaluate \llm{}s' performance when it comes to generating learning topics for specific courses. The experiment was conducted across over 100 courses and nearly 7,000 \youtube{} playlists in various subject areas. Our results indicate that \gptfour{} can produce more accurate topics for the given courses than extracted topics from \youtube{} video playlists in terms of \bert{}.
\end{abstract}

\section{Introduction \& Background}
The growth of \textit{\onlinelearning{}} has significantly impacted educational institutions and knowledge acquisition due to
its ability to facilitate continual skill development, especially in the post-pandemic world \cite{Greenhow_Graham_Koehler_2022}.
Nevertheless, online learning faces challenges in providing high-quality up-to-date educational contents \cite{Dang_Khanra_Kagzi_2022}. To tackle these issues, Artificial Intelligence (AI) can be used to automate the process of developing and maintaining high-quality up-to-date curricula \cite{tavakoli2022ai}. For instance, \cite{molavi2020extracting} used the Latent Dirichlet Allocation (LDA) algorithm to extract topics that must be covered for learning, based on the text of educational resources. While the developed methods are effective, they face challenges in terms of scalability (they only focus on limited educational areas) due to their reliance on heuristics or heavy computation \cite{Benmesbah_Lamia_Hafidi_2023}.

The introduction of promising \llm{}s, such as \textit{ChatGPT} \cite{Introducing_ChatGPT}, allows us to query larger resource databases to improve the existing automatic curriculum development algorithms. Although the opportunities \llm{}s bring to education are significant, it is crucial to consider the associated risks such as generating low-quality or factually wrong text \cite{Kasneci_Sesler_Kuchemann_Bannert_Dementieva_Fischer_Gasser_Groh_Gunnemann_Hullermeier_2023}. Therefore, we propose a novel evaluation framework to assess the performance of \llm{}s in the curriculum development context, specifically in the task of answering the question: "What topics need to be covered (both as teachers and learners) to gain required up-to-date knowledge in a course?" 

To do that, we evaluated \llm{}s efficacy to generate relevant topics on the sole basis of the course title, comparing to the baseline method derived from \youtube{} data across several educational areas. Our methodology is summarized as 1. Definition of target educational courses, 2. Constructing our dataset by querying \youtube{} as a major learning platform covering a wide range of educational areas \cite{Pires_Masanet_Tomasena_Scolari_2022}, 3. Generation of learning topics using \llm{}s (i.e., \gptthree{} and \gptfour{}), and 4. evaluating the \llm{}-generated learning topics against \youtube{} alternatives. Therefore, the key contribution of this paper is a structured, empirical methodology to understand how \llm{}s can assist curriculum developers by automatically generating relevant up-to-date topics for courses. In addition, we publish the collected dataset and \llm{}-generated topics to enable further investigations and make our approach reproducible.

\section{Method}
In this section, we explain the details of our method and data collection pipeline using \gptfour{}, \gptthree{} and \youtube{}. \Cref{fig:summary} illustrates the proposed framework to assess the effectiveness of \gptfour{} and \gptthree{} in our defined task for multiple areas. The proposed framework includes the following steps:
\begin{figure}[h]
    \centering
    \includegraphics[scale=0.65]{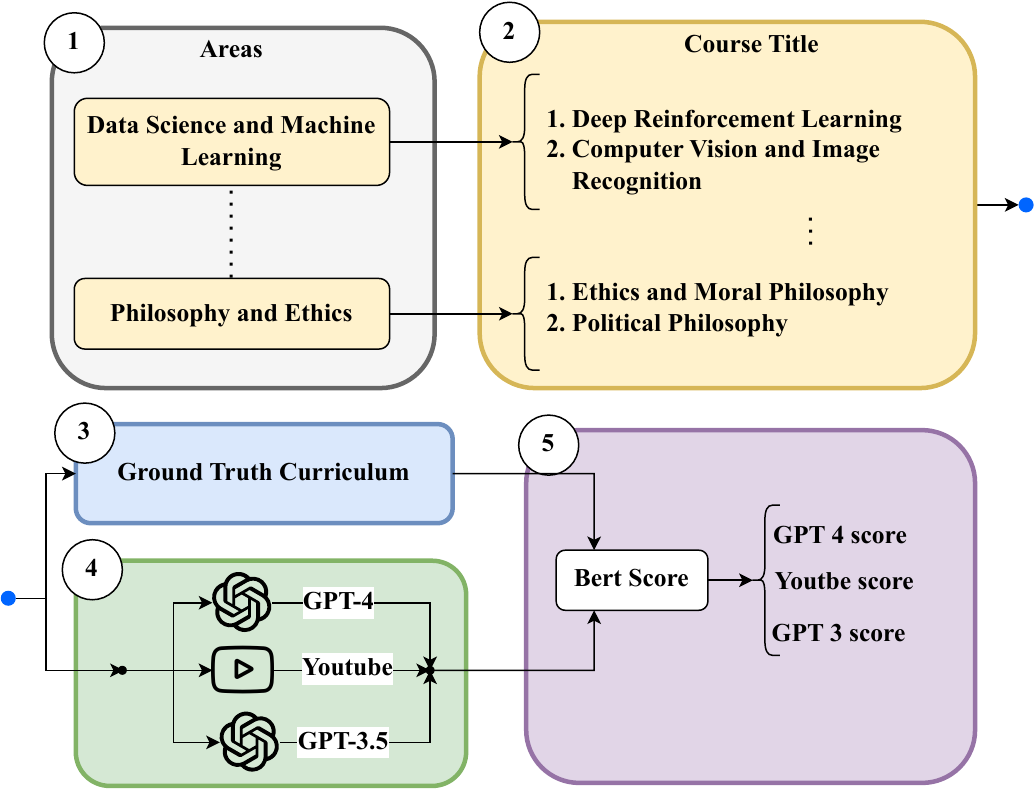}
    \caption{Data collection and evaluation pipeline: Ground truth building is illustrated in Parts 1-3; Part 4 shows the topic generation with \gptfour{}, \gptthree{}, and \youtube{}. Finally, model performance is assessed using \bert{} in Part 5.}
    \label{fig:summary}
\end{figure}
\begin{enumerate}[label=(\alph*)]
    \item \textbf{Compilation of education learning areas.} We constructed the list of target learning areas by prompting \gptfour{} to compile a list of 25 educational areas, maximizing coverage across fields to avoid bias towards certain subjects like engineering (\Cref{fig:summary}, part 1).
    \item \textbf{Course selection.} For each generated learning area, \gptfour{} proposed 4 course titles (\Cref{fig:summary}, part 2).
    \item \textbf{Quality check.} The data generated in the preceding steps was verified by four human evaluators to ensure its validity. 
    \item \textbf{\youtube{} playlist extraction.} For each course title from the previous step, we searched \youtube{} and extracted the top 10 recommended playlists. We observed that some of the playlists did not include informative titles for their videos. For instance, a few playlists included titles like \textit{lecture 1, lecture 2, etc}. These playlists were removed from the dataset.
    \item \textbf{Building ground truth dataset.} The list of video titles in each playlist served as ground truth data (i.e., covered topics) as they should represent topics in the course/playlist (\Cref{fig:summary}, part 3).
    \item \textbf{The baseline learning topics.} Our objective was to establish baselines for mapping courses with the learning topics they need to cover. We hypothesized that in the \groundcourse{}, authors searched for guidance from related courses (i.e., other \youtube{} playlists) when determining the learning topics for their courses. Therefore, we queried \youtube{} for each ground truth course title to gather the top 3 related \youtube{} playlists as \relatedcourse{} (\Cref{fig:summary}, part 4). The collected playlists that were not aligned with the title of the \groundcourse{} were removed in this step. Furthermore, duplicated phrases were frequently observed within video titles of several playlists. Since exact repetitions do not constitute distinct learning topics and are redundant, they were filtered out to be aligned with the main objective of the collected dataset. For example, the playlist titles were repeated across all videos in a number of playlists. Therefore, n-grams with n$\geq$3 were considered to identify and remove such duplicated phrases from video titles.
    \item \textbf{Generating \llm{} titles.}  \gptfour{} and \gptthree{} were prompted with the same course titles (ground truth course titles) to generate AI-produced candidate topics (\Cref{fig:summary}, part 4). Both \gptfour{} and \gptthree{} were sampled 3 times in order to control for uncertainty. This simulated the process of utilizing \llm{}s to develop an automatic curriculum (list of learning topics) for a course.  As an example, our prompt for \emph{Political Philosophy} was:
          \begin{itemize}
              \item[] \textit{I am preparing to teach a course titled 'Political Philosophy'. Could you please generate a list of topics suitable for teaching this course? The topics should be in English and each one should be listed on a separate line.}
          \end{itemize}

    \item \textbf{Evaluation.} The \emph{\textbf{BERTScore}} \cite{Zhang_Kishore_Wu_Weinberger_Artzi_2020} was used to evaluate the performance of \llm{} generated curricula (learning topics) in comparison with the \youtube{} baseline curriculum (\Cref{fig:summary}, part 5). One of the key advantages of \bert{} is its use of contextual embedding, which makes it robust to syntactic differences \cite{Zhang_Kishore_Wu_Weinberger_Artzi_2020}. For the comparison, the cosine similarity between the candidate embeddings (\llm{} generated curriculum) and the reference embeddings (\youtube{} baseline) was computed. The Bert-related computations were done by using the \texttt{bert-score} python package\footnote{\url{https://pypi.org/project/bert-score/}} with a \texttt{xlnet-base-cased} model.
\end{enumerate}
Applying the above-mentioned steps produced a dataset containing 727 ground truth playlists, 6631 \relatedcourse{}, and 6 (i.e., 3 for \gptfour{} and 3 for \gptthree{}) sets of LLM-generated learning topics per ground truth playlist\footnote{The final dataset can be found from: \href{https://bit.ly/raw_dataset}{\texttt{https://bit.ly/raw\_dataset}}}.

\section{Results and Discussion}
\begin{figure*}[t]
    % \hspace{-2cm}
    \resizebox{\textwidth}{!}{\input{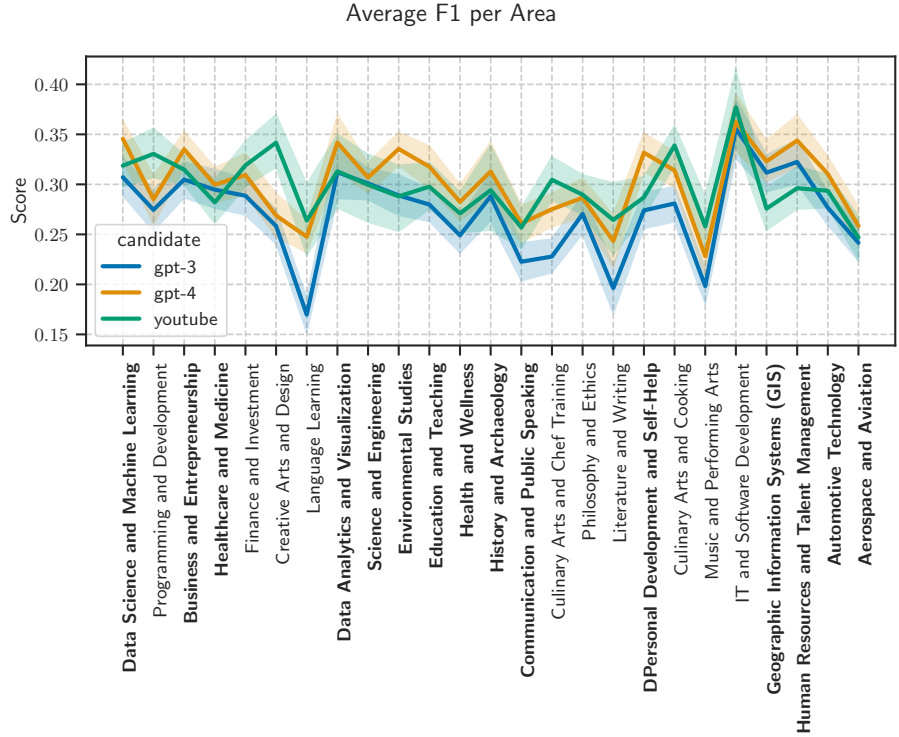}}
    \caption{Performance for each \textbf{area}. The graph shows the mean and 95\% confidence interval. Areas where \gptfour{} \textbf{outperformed} \youtube{} are bold faced.}
    \label{fig:perf}
\end{figure*}

We compared the performance of \gptthree{} and \gptfour{} generated learning topics to \youtube{} video titles (in \relatedcourse{}) using \bert{} $F_1$. \Cref{tab:perf_avg_model} summarizes the average and standard deviation of \bert{} for each method. \gptfour{} achieved slightly higher with an $F_1$ score of $0.30 \pm 0.10$ than \youtube{} with a score of $0.29 \pm 0.11$. \gptthree{} performed worse than the other candidate models, which was expected since \gptthree{} is a smaller model than \gptfour. Nevertheless, both GPT models have better precision than \youtube{}. Essentially, this means that when GPT models generate topics, more of their topics are present in the ground truth dataset, compared to the baseline. But at the same time, \gptfour{} misses some extra topics, which are relevant to the ground truth (lower recall). Finally, we analyzed the performance of the models based on each area as illustrated in \Cref{fig:perf}. This confirms that, in general, \gptfour{} is performing better than \youtube{} and \gptthree{}\footnote{Detailed performance metric per area can be found from: \href{https://bit.ly/topic_model_performance}{\texttt{https://bit.ly/topic\_model\_performance}}}.
\setlength{\tabcolsep}{12pt}
\begin{table}[h]
    \sisetup{
  table-align-uncertainty=true,
  separate-uncertainty=true,
}
%% local redefinitions
\renewrobustcmd{\bfseries}{\fontseries{b}\selectfont}
\renewrobustcmd{\boldmath}{}
    \centering
    \caption{Average and Standard Deviation of \bert{} for candidate models. }
    \label{tab:perf_avg_model}
    \begin{tabular}{lllll}
    
        \toprule
                         & $F_1$                & Precision       & Recall            \\
        Candidate        &                   &                 &                 & \\
        \midrule
        \gptthree{}   & 0.27 $\pm$ {0.10} & 0.30 $\pm$ 0.12 & 0.25 $\pm$ 0.10   \\
        \gptfour{}   & \textbf{0.30} $\pm$ \textbf{0.10} &  \textbf{0.32}  $\pm$  \textbf{0.12} & 0.29 $\pm$ 0.10   \\
        \youtube{} & 0.29   $\pm$ 0.11 & 0.29 $\pm$ 0.13 & \textbf{0.30} $\pm$ \textbf{0.13}   \\
        \bottomrule
    \end{tabular}
\end{table}

Although our results are promising, some limitations should be considered. We acknowledge that considering more learning areas will provide a better estimation of the performance of \llm{}s in our target task. Also, the results are limited to using \youtube{} as ground truth, and not necessarily generalizable to other large repositories or resources of learning content. However, this is not the limitation of the evaluation framework, but the dataset we constructed. It should be noted that \youtube{} is a powerful opponent in our defined task as it uses more information (e.g., video content and description) beyond just the video title to offer related playlists (courses). Therefore, despite the potential information discrepancy, the \llm{}s achieved competing results and this shows the potential of these \llm{}s when it comes to generating relevant topics for a course.

\section{Conclusion and Future Work}
In this study, we present a framework and comprehensive dataset to explore LLMs' capability in curriculum development, particularly in generating relevant learning topics. Our defined task was finding the learning topics that should be covered in a course. In 25 different learning areas and their including courses, \gptfour{} models performed reasonably well in terms of $F_1$ score which highlighted the benefits of such models for curriculum development.

We recognize that certain decisions in this study, such as how we define educational areas, might have valid alternatives. However, our focus was on presenting our framework as a case study. We aimed to demonstrate how YouTube (as a popular learning platform) can serve as an indicator for outdated curricula, and how Large Language Models (LLMs) can offer a quick, effective update method in these situations. Further research, involving more diverse data sources, such as Coursera or university curricula, could improve the proposed dataset and refine our framework. Also, further investigations need to be carried out on prompt engineering (for example to provide more context) and hyperparameter tuning to investigate whether we can improve the LLMs' performance.

\bibliographystyle{splncs04}
\bibliography{paper}
\end{document}